\documentstyle[12pt]{article}
\begin{document}
\title{
Quantum Intermittency in
Almost-Periodic Lattice Systems Derived from their Spectral
Properties}
\author{Giorgio Mantica \\
Istituto di Scienze Matematiche, Universit\`a di Milano a Como, \\
Via Lucini 3, I-22100 COMO, ITALY \\
{\sc mantica@mi.infn.it}\\
}
\date{ }
\maketitle
\begin{center}
{\em This paper is dedicated to the memory of Professor Joseph
Ford,}\\ 
{\em teacher, original researcher, and founder of Physica D}\\
\end{center}
\begin{abstract}
Hamiltonian tridiagonal matrices characterized by
multi-fractal spectral measures in the family of
Iterated Function Systems
can be constructed by a recursive technique here described. 
We prove that
these Hamiltonians are almost-periodic.
They are suited to describe
quantum lattice systems
with nearest neighbours coupling,
as well as chains of linear classical oscillators,
and electrical transmission lines.

We investigate numerically and theoretically
the time dynamics of the systems so constructed.
We derive a relation linking 
the long-time, power-law behaviour 
of the moments of the position operator, expressed by a
scaling
function $\beta$ of the moment order $\alpha$, and
spectral multi-fractal dimensions, $D_q$, via 
$
          \beta(\alpha) = D_{1-\alpha}.
$
We show cases in which this relation is exact, and cases 
where it is only approximate, unveiling the reasons for
the discrepancies.
\end{abstract}

\section{Introduction}

Usually, the study of almost/quasi-periodic systems starts by assigning
a suitable
rule for building a quantum Hamiltonian operator,
and then proceeds to the determination
of its spectral quantities \cite{magn}
(which is frequently a hard task) and of the time dynamics it generates.
In so doing, {\em multi-fractal} energy spectra have been 
frequently observed, and anomalous characteristics
of the time evolution have been exhibited
\cite{koh,yam,geis1,ig1,igm}.

These findings raise the question if multi-fractal spectra
are typical in almost/quasi-periodic systems \cite{gen}, and, 
vice versa, if
almost/quasi-periodicity is always associated with singular continuous
spectral measures. The relations between this
pair (spectral multi-fractality and Hamiltonian
almost-periodicity) and
the time dynamics generated via Schr\"odinger's equation 
are also interesting, and intricate: do the former 
always imply anomalous scaling relations of physical
quantities like, for instance, the expectation value of the
position operator ? Can we make any quantitative statement to
this effect ?

In this paper, we employ a new algorithm
for deriving a Hamiltonian operator (called a {\em Jacobi
matrix} because of its mathematical nature) with a pre-assigned 
spectral measure in the vast class of {\em Iterated Function
Systems} (I.F.S.). This technique provides us with an ideal
patient for our surgical table, who can be fully dissected
and analyzed. 
In particular, we provide evidence that
Jacobi matrices associated with I.F.S. are
almost periodic, and we argue that this is
likely to be the typical case in a large class of measures
with fractal support.

The analysis which can be carried out in this example
permits us to compute exactly the asymptotic
behaviour of the wave--function projections, for short and long times.
By introducing a renormalization approach in 
the theory of orthogonal polynomials, 
we also derive a relation linking 
the asymptotic power-law growth of the moments of the
position operator and multi-fractal generalized dimensions.
This theory explains the phenomenon that we have termed
{\em quantum intermittency}.

The specific properties of I.F.S. are crucial
for our theory; Yet, since
particular I.F.S.'s can be found so to
approximate arbitrarily well (in a technical sense)
any ``fractal'' measure \cite{gio2,gio,vrs,gio-guzzi}, the results obtained 
in the I.F.S. class may have a much wider generality.

We shall present our results
as follows: in Section II we introduce the 
general formalism of I.F.S. and
of Jacobi matrices, employed to solve the {\em inverse
problem} of finding an Hamiltonian with a given spectrum.
This formalism is then applied in Sect. III to derive
a stable solution algorithm.
The almost-periodic properties of the Hamiltonian so determined
are studied numerically in Sect. IV, and 
the intermittent quantum dynamics it generates is then discussed in 
Sect. V and VI.
The Conclusions
summarize the work and present some previews on further investigations.

\section{I.F.S. and their Jacobi Matrices}

Systems of linear 
iterated functions \cite{hut,dia,dem,ba2} are finite
collections of maps 
\begin{equation}
\label{mappi}
    \phi_{i} (x) := \delta_{i} x + \beta_{i}, \;\;  i = 1, \ldots, M ,  
\end{equation}
where $\delta_{i},\beta_{i}$ are real constants, and 
where the contraction rates 
$\delta_{i}$ have modulus less than one.
Without loss of generality, we may assume that 
each $\phi_i$ maps $[0,1]$ into itself, and that $\phi_1(0) = 0$.

A probability, $\pi_i$, is
associated with each map:
$\pi_{i} > 0$, $\sum_{i} \pi_{i} = 1$. 
Employing these probabilities,
a measure over $[0,1]$ can be defined 
as the unique positive
measure satisfying the balance property 
\begin{equation}
\label{bala} 
   \int _{0}^1 f \; d\mu \; = 
   \sum_{i=1}^{M}
   \; \pi_{i}
   \; \int _{0}^1  \; 
 (f \circ \phi_{i}) \; d\mu ,  
\end{equation}
for any continuous function $f$.
This measure is supported on $A$,
the subset of $[0,1]$ which solves
the equation 
\begin{equation}
\label{attra}
    A=\bigcup_{i=1,\ldots ,M}\;\phi_i(A),
\end{equation}
The set $A$ is invariant under
the action of shrinking it to smaller copies of itself, and glueing 
them together. Because of 
eq. (\ref{attra}), the geometry of this set is typically fractal
(except for special choices of the map parameters); 
In turn,
the balance relation (\ref{bala}) is responsible for the
multi-fractal properties of the measure $\mu$.
In fact, let us consider a
{\em disconnected} I.F.S., that is to say, one for which
the sets $\phi_i(A)$ do not intersect each other. Under these
circumstances,
the multi-fractal properties of the balanced measure are easily
computable: the spectrum of generalized dimensions
$D_q$ follows from the equation
\begin{equation}
     \sum_{j=1}^{M} 
     \pi_j^q \delta_j^{-\tau} = 1 ,
\label{multi1}
\end{equation}
whose unique real solution defines $\tau$ as a function of $q$,
and leads to $D_q = \frac{\tau(q)}{q-1}$. In virtue of
this relation, one can tune the map parameters 
to obtain various multi-fractal spectra.

The problem of determining a Hamiltonian
possessing $\mu$ as spectral measure can be solved
\cite{kac} considering
the set of associated orthonormal polynomials,
$\{p_n\}$:
   \begin{equation}
    \int p_i(x) p_k(x) \; d\mu(x) = \delta_{i,k}.  
   \end{equation}
In fact, any such set of polynomials is 
characterized by a
three-terms recurrence relation which can be written
\begin{equation}
\label{nor2} 
   x p_j (x) = r_{j+1} p_{j+1}(x) + A_j p_j(x)  + r_{j} p_{j-1}(x),
\end{equation}
or, in matrix form
\begin{equation}
\label{jac2} 
    H p (x) = x p (x).
\end{equation}
In the above,
$p(x)$ is the vector whose components are the orthonormal polynomials
evaluated at site $x$, and $H$ is the Jacobi matrix, which is
constructed as the real, symmetric, tridiagonal
matrix whose diagonal and outer diagonals are the vectors $A_j$ and $r_j$,
respectively: 
\begin{equation}
\label{jac11} 
   H_{i,i} = A_i , \; \; \;  H_{i+1,i} = H_{i,i+1} = r_{i+1} , \;
\; i = 0,1,\ldots .  
\end{equation}
$H$ 
defines a nearest neighbours lattice system, with site energies
$A_i$ and hopping constants $r_i$.
Similarly, $H$ can describe a linear array of masses coupled 
by springs, and also an electrical transmission line,
whose characteristics vary from one element to the next.

Standard theory proves that, letting the Jacobi matrix
$H$ act in $l_2$
(the space of square summable sequences,
whose canonical basis will be indicated
by $\{e_0,e_1,\ldots\}$) 
the spectral measure of $H$ 
with respect to the vector $e_0$ (the {\em local density 
of states}
of physical jargon) is precisely $\mu$: in fact, one has
\begin{equation}
\label{jac8} 
	(e_0, g(H) e_0) = \int g(x) \; d\mu(x),
\end{equation}
for well-behaved functions $g$.
This is the 
theoretical solution of the inverse problem we have proposed.
In order to translate it into a {\em practical} solution,
we need to compute the Jacobi matrix coefficients
starting from the measure $\mu$, i.e.
from the map parameters defining the I.F.S. 

\section{A Stable Technique for Computing I.F.S. Jacobi Matrices}

The problem of constructing the Jacobi matrix associated
with I.F.S. measures is hard, 
and the usual techniques of {\em polynomial sampling} 
\cite{gax,steve,gaut} 
are plagued by exponentially
increasing errors which allow only computation of very few 
Jacobi matrix coefficients \cite{cap}.
Alternatively,
the sole technique available so far has been
an {\em algebraic} procedure
programmed in {\em MAPLE} by Vrscay \cite{vrs2}. Yet, 
it is severely limited by memory and time requirements.
To overcome these difficulties 
we have devised a direct algorithm applicable to I.F.S. 
measures.

We first observe that, for any $n$, 
\begin{equation}
\label{le1}
   p_n(\phi_i(x))=\sum_{l=0}^n\Gamma _{i,l}^n\;p_l(x),
   \;\;\; i = 1,\ldots, M.
\end{equation}
This is immediate, since 
$p_n(\phi_i(x))$ is an $n$-th degree polynomials which 
can be expanded on the first $n$ orthogonal polynomials.
Less immediate is to derive a recursive rule for
the coefficients $\Gamma _{i,l}^n$, $l=0,\ldots ,n$.
It turns out that, at fixed $n$, they can be determined
from the map parameters,
and from the Jacobi matrix entries $A_j$, for $j=0,1,\ldots ,n-1$, and 
$r_m$, for $m=0,1,\ldots ,n$. 
In fact, (dropping
for simplicity the map index $i$) we have that
$p_0(\delta x+\beta )=p_0(x)$, and hence $\Gamma _0^0=1$. 
Suppose now that $\Gamma^{k}_l$ is known for $k=0,\ldots,n-1$
and all relative $l$'s:
from eq. (\ref{nor2}) we obtain
the complete decomposition of
$p_n(\delta x + \beta)$ over $p_l$, $l=1,\ldots ,n$:
\begin{equation}
\label{rec4}
   \begin{array}{ll}
   r_np_n(\delta x+\beta )
   =& (\beta -A_{n-1})\sum_{l=0}^{n-1}\Gamma_l^{n-1}p_l(x) + \\
  &  +\delta \sum_{l=0}^{n-1}\Gamma_l^{n-1}(r_{l+1}p_{l+1}(x)+A_lp_l(x) 
   +r_lp_{l-1}(x)) + \\
   & -r_{n-1}\sum_{l=0}^{n-2}   \Gamma _l^{n-2}p_l(x). \\
 \end{array}
\end{equation}
Equation (\ref{rec4}) allows now the determination of
the coefficients $\Gamma _l^n$. 

We observe that the highest order polynomial, $p_n$,
appears twice in the above equation,
always in the form of the product $r_n p_n$: hence, the
coefficients in the expansion of 
the polynomial $r_n p_n$ can be determined {\em without}
knowing $r_n$. Therefore, if we let $\tilde{p}_n(x) = r_n p_n(x)$,
a second decomposition can be written as
\begin{equation}
\label{le1b}
   \tilde{p}_n(\phi_i(x)) = \tilde{\Gamma}^n_{i,n} \; \tilde{p}_n (x)
+  \sum_{l=0}^{n-1} \tilde{\Gamma}^n_{i,l} \; p_l(x) ,  
\end{equation}
where the coefficients $\tilde{\Gamma}$ can be computed recursively 
from eq. (\ref{rec4}),
on the basis of the knowledge of only $A_j$, $r_j$, for 
$j=0,1,\ldots,n-1$. 

We can now compute the
non-diagonal entries of the Jacobi matrix:
from eq. (\ref{nor2}) we write 
\begin{equation}
\label{rec8} 
   r_n^2 = \int \tilde{p}_n(x) x p_{n-1}(x) \; d\mu .  
\end{equation}
Hence, using the balance property (\ref{bala}) and 
eqs. (\ref{le1}),(\ref{le1b}) this becomes
\begin{equation}
\label{rec9}
   r_n^2 = \sum_{i=1}^M \pi_i \int (\delta_i x + \beta_i) [
\sum_{m=0}^{n-1} \sum_{l=0}^{n-1} \tilde{\Gamma}^n_{i,m} \Gamma^{n-1}_{i,l}
p_m(x) p_l(x) + \sum_{l=0}^{n-1} \tilde{\Gamma}^n_{i,n} \Gamma^{n-1}_{i,l} 
\tilde{p}_n(x) p_l(x) ] d\mu .  
\end{equation}
Again, we can use the 
recurrence relations (\ref{nor2}), to get 
\begin{equation}
\label{rec10}
   r_n^2 = \sum_{i=1}^M \pi_i \; (B_i + C_i + D_i) ,  
\end{equation}
where we have put: 
\begin{equation}
\label{rec11}
   B_i = \sum_{l=0}^{n-1} (\beta_i + \delta_i A_l) 
   \tilde{\Gamma}^n_{i,l} \Gamma^{n-1}_{i,l} ,  
\end{equation}
\begin{equation}
\label{rec12}
   C_i = \delta_i \sum_{l=0}^{n-2} r_{l+1} ( \tilde{\Gamma}^n_{i,l} 
   \Gamma^{n-1}_{i,l+1} + \tilde{\Gamma}^n_{i,l+1} \Gamma^{n-1}_{i,l}) ,  
\end{equation}
and 
\begin{equation}
\label{rec13}
   D_i = \delta_i \tilde{\Gamma}^n_{i,n} \Gamma^{n-1}_{i,n-1}
r_n^2 .  
\end{equation}
Because of contractivity of the maps, $|D_i|r_n^{-2} < 1$. 
Therefore,
$r_n^2$ (and hence  $r_n>0$) can be
computed from eq. (\ref{rec10}),
on the basis of the knowledge of
the coefficients in the
expansions (\ref{le1}) of order $n-1$, 
of order $n$ in (\ref{le1b}), of the map
parameters, 
and of the matrix entries $A_j$, $r_j$, for $j=0,1,\ldots ,n-1$. 

A similar trick allows the computation of the
diagonal entries $A_n$;
We use eqs. (\ref{bala}) and (\ref{nor2})
(integrals are taken with respect to $\mu$):
\begin{equation}
\label{rec6}
   A_n = \int x p^2_n(x) = \sum_{i=1}^M \pi_i \int (\delta_i x +
\beta_i) p_n^2(\delta_i x + \beta_i) \; = \; \sum_{i=1}^M \pi_i \int
(\delta_i x + \beta_i) \sum_{m,l=0}^{n} \Gamma^n_{i,l} \Gamma^n_{i,m} p_l(x)
p_m(x) .  
\end{equation}
Using the orthonormality properties of the sequence $p_n$, and 
the recurrence relation, eq. (\ref{nor2}),
we get
\begin{equation}
\label{rec7} 
   A_n = \sum_{i=1}^M \pi_i [ \sum_{m=0}^{n} (\Gamma^n_{i,m})^2 \;
(\beta_i + \delta_i A_m) + \sum_{m=0}^{n-1} \Gamma^n_{i,m} \Gamma^n_{i,m+1}
\delta_i (r_m+r_{m+1}) ],  
\end{equation}
thereby determining $A_n$ as a function of
the coefficients in eq. (\ref{le1}) of order $n$ fixed,
of the map parameters,
and of the matrix entries $A_j$,
for $j=0,1,\ldots ,n-1$, and $r_m$, for $m=0,1,\ldots ,n$.

These results can be properly
chained into an iterative
construction of the Jacobi matrix $H$:
The algorithm is structured as follows:

\begin{itemize}
\item  {\em Initialization}. 
At the first step, we have $A_0=\mu_1$, $r_0=0$,
$\Gamma^0_0=1$. The first order moment of $\mu$, $\mu_1$,
can be simply computed from eq. (\ref{bala}).
\item  {\em Iteration}. 
Suppose that  $A_l$, $r_l$,
and $\Gamma^l$  are known for $l=0,1,\ldots ,n-1$. Then
we:
\begin{itemize}
\item  {\em Compute $\tilde \Gamma ^n$}. We use 
equations (\ref{rec4} - \ref{le1b}).
\item  {\em Compute $r_n$}. We use eqs. 
(\ref{rec8} - \ref{rec13}).
\item  {\em Compute $\Gamma ^n$}. This is immediate at
this stage.
\item  {\em Compute $A_n$}. We use eqs. (\ref{rec6} - \ref{rec7}).
Then we iterate the procedure.
\end{itemize}
\end{itemize}

Graphically:
\begin{equation}
\label{recu}
  \left(
   \Gamma^{n-1}, \; \; 
\begin{array}{l}
r_0,\ldots,r_{n-1} \\ 
A_0,\ldots,A_{n-1} \\  
\end{array} \right)
\Rightarrow \tilde{\Gamma}^n \Rightarrow r_n \Rightarrow \Gamma^n
\Rightarrow A_n \Rightarrow \left( \Gamma^{n}, \; \; 
\begin{array}{l}
r_0,\ldots,r_{n} \\ 
A_0,\ldots,A_{n} \\  
\end{array} \right)
\end{equation}

In a separate work \cite{cap}
we have analyzed the reasons of the failure of
classical polynomial sampling \cite{gax,gaut}
when applied to singular measures,
and assessed the numerical stability of the
recursive algorithm presented above.
We have observed a polynomial
error propagation with respect to matrix order for the
recursive algorithm,
while using the classical algorithms
the error growth was found to be exponential.

\section{Almost Periodicity of I.F.S. Jacobi Matrices}

Having devised a stable solution of the Hamiltonian inverse
problem, we can study the properties
of large Jacobi matrices. Fig. 1 shows
an I.F.S. measure, one of its orthogonal polynomials, and the
begining of the
sequence of $r_n$ coefficients.
Let us focus our attention on the last.

We can clearly observe a zero frequency
(the average value), a $\pi$ frequency (flipping up and down),
and clearly other frequencies are present in the sequence.
A Fourier analysis is simply effected writing
\begin{equation}
    r_n = \sum_k F_k e^{i n \omega_k} .
\label{fu1}
\end{equation}
This sum may not converge in the usual sense,
and it might have to be replaced by an integral in the case of
a continuous component in the ``spectrum'' of the sequence
$r_n$.
If the continuous component is absent, the system is 
almost-periodic.
Within this case, if the set of frequencies $\omega_k$
can be derived from a finite set of periods,
the sequence $r_n$ is {\em quasi-periodic}:
that is, this is the case
if there exist suitable $\Omega_1, \ldots ,
\Omega_p$ such that for all $k$ the frequency $\omega_k$ can be written
$\omega_k = n_1 \Omega_1 + \ldots + n_p \Omega_p$,
for integer $n_1,\ldots,n_p$.

A numerical, fast Fourier analysis of the sequence $r_n$
is presented in Fig. 5,
where peaks in the distribution of
$|F_k|^2$ with a clear hierarchical structure are 
observed. These peaks seem to suggest the presence 
of a point component in the spectrum of this sequence.
Yet, care has always to be exerted to assess 
this fact numerically.
To obtain a further piece of evidence
we performed an analysis of the phase 
of $F_k$ around these peaks, like that shown in
Fig. 2, and found a $\pi$ discontinuity, which 
indicates \cite{andrei} that they are indeed
related to a point component. The sequence $r_n$ is therefore
almost periodic. 

Since no simple rational relation
among the peak sequences seems to hold, numerical evidence 
seems to suggest that the sequence is not quasi-periodic.
Our numerical investigations have shown that
these characteristics are typical in the class of
Hamiltonian associated with I.F.S. measures, supported on Cantor
sets.
In view of the approximation properties of I.F.S. measures,
this result is likely to be much more general:
indeed, in the family of Jacobi matrices associated with
real Julia sets \cite{danbel},\cite{barn}, 
which can be well approximated by I.F.S.,
limit periodicity of the sequence 
$r_n$ has been proven directly \cite{danbak}.
The problem of a formal proof is therefore open.  

\section{Quantum Dynamics of Almost Periodic Lattice Systems}

Jacobi matrices generate a
quantum dynamics in $l_2$ 
via Schr\"odinger's
equation, 
\begin{equation}
   i \frac{d \psi}{d t}  = H \psi, \;\;
   \psi(0) = e_0 := (1,0,\ldots) .
\label{ev1}
\end{equation}
The initial state of the evolution, $e_0$, is
the zeroth lattice state. In oscillator terms, this 
corresponds to a situation where the first mass is displaced
from its equilibrium position, while all the other masses are
at rest in their equilibria. In electrical terms, the current (or
the voltage) is
non-zero only in the first element of the transmission line described
by the Jacobi matrix $H$.

The solution of Schr\"odinger equation can be formally
obtained as \cite{dan,parl}
\begin{equation}
   c_n(t) := (e_n, e^{-itH} e_0)  = \int
   e^{-itx} p_n(x) \; d\mu(x) ,
\label{crux}
\end{equation}
where $c_n(t)$ is the component of $\psi(t)$ at the
$n$-th lattice state. Equation (\ref{crux}) shows that this
component is 
the Fourier transform of the orthogonal polynomial $p_n$
with respect to the spectral measure $\mu$.
This fact allows us to derive important results.

Firstly,
the asymptotic behaviour for small $t$ can be controlled as follows:
$|c_n(t)|^2 \sim t^{2n}$. In fact,
\begin{equation}
     c_n(t) =
     \int d\mu(x) \; p_n(x) 
	 \sum_{l=0}^{\infty} \frac{(-it)^l}{l!}  x^l =
	 \sum_{l=n}^{\infty} \frac{(-it)^l}{l!} \int d\mu(x) p_n(x) x^l .
\label{rev10}
\end{equation}
Because of the orthogonality properties of the set $p_n$ 
this expansion begins with $l=n$, which proves the result.

Secondly, in the infinite time limit, 
denoting by $\overline{S}_n (T)$ the  
time average of $|c_n|^2$ up to time $T$,
\[
\overline{S}_n(T) 
= \frac{1}{2T} \int_{-T}^{T} |c_n|^2 (t) \; dt ,
\]
we have that
\begin{equation}
\overline{S}_n(T) 
\sim T^{-D_2}
\label{decay}
\end{equation}
for all $n$, a result which involves the correlation dimension  $D_2$
of the fractal measure $\mu$.
The case with $n=0$ is implicitly contained in Bessis 
{\em et al.} \cite{turca}, and was 
originally proposed in the present context by Ketzmerick 
{\em et al.} \cite{geis1}. Successively,
it has attracted a lot of attention,
mainly from cultors of mathematical rigour.
Our generalization has the advantage of requiring a simple
proof, via the usage of the Mellin transform, as in
\cite{turca}.
In fact, we write
\begin{equation}
\overline{S}_n(T) =
      \int d\mu(x) \int d\mu(y) \; 
	  \frac{\sin \; (x-y)T}{(x-y)T} \;
	    p_n(x) p_n(y) .
\label{rev8}
\end{equation}
To find the asymptotic behaviour of eq. (\ref{rev8}),
we take the Mellin transform, $M_n(z)$, of 
$\overline{S}_n(T) $:
\begin{equation}
   M_n(z) = \int T^{z-1} \overline{S}_n(T) dT = G \times
      \int d\mu(x) \int d\mu(y) \; 
	  \frac{p_n(x) p_n(y)}
	  {|x-y|^{z}} = G(z) \times E_n(z) ,
\label{rev9}
\end{equation}
where $G(z) = \Gamma(z-1) \sin (\frac{\pi}{2}(z-1))$,
and where $E_n(z)$ is defined implicitly by the last equality.
The dominating power law in the
long time behaviour of $S_n$ is determined by the 
divergence abscissa of $M_n(z)$: that is to say, 
$\overline{S}_n(T) \sim T^{-w}$, where
$w$ is the
largest real $z$ for which $M_n(z)$, hence $E_n(z)$
converges. 
It is apparent from eq. (\ref{rev9}) that
the divergence of $E_n$ is piloted by the
small scale structure of the measure $\mu$.
Because the polynomials $p_n$ are smooth functions,
with bounded derivatives on the support of $\mu$,
the divergence abscissa of $E_n$ is the same for all $n$,
and, in particular, it coincides with that of $E_0$.
$E_0(z)$ is known as
the {\em generalized electrostatic energy}
of the measure $\mu$ and its divergence abscissa
is known to be $D_2$ \cite{turca}, the correlation dimension 
of the measure $\mu$.

It is important to remark that the domains of validity
of the asymptotic expansions just derived 
are not uniform in $n$. 
This adds to the difficulty of the problem to be discussed 
in the next Section.

\section{Renormalization Theory of Quantum Intermittency}

An important characteristics of the quantum motion introduced
in the previous section is the way
it spreads over the $l_2$ lattice basis, $\{e_n\}$.
In fact, in oscillator terms, spreading corresponds to energy transmission 
along the linear chain, be it mechanical or electrical.
In quantum mechanical terms, it corresponds to 
unbounded motion of the lattice particle, of the kind 
treated only qualitatively by R.A.G.E. theorems.
To gauge this phenomenon, we define 
the moments of the position operator $\hat{n}$: 
\begin{equation}
\nu_\alpha(t) := (\psi(t), \hat{n}^\alpha
\psi(t)) = \sum_n n^\alpha |c_n(t)|^2. 
\label{galas}
\end{equation}
Their asymptotic behaviour follows a power law,
\begin{equation}
\nu_\alpha(t) \sim t^{\alpha \beta},
\label{sca1}
\end{equation}
where $\beta$ is a non-trivial function of the
moment order $\alpha$.
In \cite{igm,parl}
we found that $\beta$ is convex, non-decreasing, and
non-constant even in the case of a one-scale Cantor set,
characterized by trivial thermodynamics:
this is what we call {\em quantum intermittency}.
Corrections to eq. (\ref{sca1})
can also be observed in the form
of log-periodic oscillations of $\nu_\alpha(t)$, super-imposed
to its leading behavior. They can be explained by the 
Mellin-type analysis presented in the previous section.

We can estimate the function 
$\beta(\alpha)$ on the basis of simple 
renormalization group considerations.
For simplicity,
let us consider an I.F.S. with $M$ maps, of equal
probability
$\pi_i=\frac{1}{M}$. Let this I.F.S. be non-overlapping.
Then, let $I$ be the smallest interval containing
$A$, the I.F.S. attractor, and let $I_l$ be the image of 
$I$ under the map $\phi_l$. Clearly,
$I_l \cap I_m = \emptyset$ if $l \neq m$, and the measure
$\mu$ restricted to $I_l$ is a linearly rescaled copy of 
the original. Then, as a first approximation, we can 
assume that the orthogonal polynomials of the restricted
measure are also obtained by linear rescaling of the
original polynomials:
\begin{equation}
   p_{Mn} (\phi_l(x)) = \sum_{k=0}^{Mn} \Gamma_{l,k}^{Mn} p_k(x)
   \simeq \sigma^n_l p_n(x) ,
\label{est1}
\end{equation}
where $\sigma^n_l = \pm 1$. In other words, we assume 
a very simple form for the coefficients $\Gamma^{Mn}$,
which amounts to making a renormalization ansatz.

Let us now consider $\overline{S}_{Mn}(T)$, as defined above.
Because of the balance property (\ref{bala}), it can be written
\begin{equation}
\overline{S}_{Mn}(T) =
\sum_{l,m=1}^{M}
\pi_l \pi_m
 \int d\mu(x) \int d\mu(y) \;
 \frac{\sin T(\phi_l(x)-\phi_m(y))}{T(\phi_l(x)-\phi_m(y))} 
  p_{Mn}(\phi_l(x)) p_{Mn}(\phi_m(y)) .
\label{est3}
\end{equation}
In the previous equation, 
$\phi_l(x)$ and $\phi_m(y)$ belong to $I_l$ and $I_m$,
respectively. If
$l \neq m$, these intervals are separated by a finite gap.
As $T$ tends to infinity, these contributions tend to zero 
as $T^{-1}$. We can therefore retain only the diagonal 
terms in eq. (\ref{est3}). 

If we now employ the approximate estimate (\ref{est1})
in the r.h.s. of eq. (\ref{est3}) we can write
\begin{equation}
\overline{S}_{Mn}(T) =
\sum_{l=1}^{M}
\pi_l^2
\overline{S}_{n}(\delta_l T).
\label{est5}
\end{equation}
This too is a sort of renormalization equation which links the 
wave-function component at site $Mn$ and time $T$ to
the component at site $n$ and at shorter times $\delta_l T$.
When inserted in eqs. (\ref{galas}), eq. (\ref{est5})
implies that the growth exponent $\beta$ associated 
with the averaged moments $\overline{\nu}_\alpha$ via eq. (\ref{sca1})
must satisfy the relation
\begin{equation}
    1 = M^{\alpha-1} \; \sum_{l=1}^{M} \delta_l^{\alpha \beta}.
\label{est9}
\end{equation}
Comparing this result with eq. (\ref{multi1}) we obtain
the crucial equation
\begin{equation}
    \beta (\alpha) = D_{1-\alpha} ,
\label{multap}
\end{equation}
which links multi-fractal properties and time dynamics.
In particular, eq. (\ref{multap}) 
implies that $\beta(0) = D_1$, which
is consistent with the rigorous result 
$\beta(0) \geq D_1$ \cite{ig1}. Notice that $\beta(0)$ can be
defined by a limiting procedure on
$\beta(\alpha)$, or by the evolution
of the logarithmic moment. We have also $\beta(1) = D_0$.

Because of the rough approximation involved in eq. (\ref{est1}),
and because for the validity of eq. (\ref{est5}) 
both $c_{Mn}$ and $c_n$ need to be in their asymptotic regimes, 
we do not expect eq. (\ref{multap}) to be 
always exact.
Indeed, in Fig. 4 we have considered a family of 
I.F.S. measures, characterized by $M=2$, 
$\delta_2 = \frac{2}{5}$,
$\beta_1=0$,
$\beta_2=\frac{3}{5}$,
$\pi_1=\frac{3}{5}$, and
$\pi_2=\frac{2}{5}$.
The contraction rate $\delta_1$ is allowed
to vary in the range $[.2,.5]$, which implies a significant variation
both in the structure of the support of the balanced measure and
in its multi-fractal properties.
Plotted in Fig. 4 are the scaling exponents $\beta(0)$
and $\beta(1)$, compared with the multi-fractal dimensions
$D_1$ and $D_0$, respectively.
We observe a substantial agreement between the two data sets,
dynamical and multi-fractal.
Numerically,
the discrepancy is always less than five percent.
We can therefore conclude that the relation 
(\ref{multap}) catches some essential part of the
physics. Yet, the situation is more complicated,
as the following pair of examples show.

The first is a magnificent 
counter-example. Let us consider a new class 
of I.F.S. measures (and related Hamiltonians)
characterized by $M=2$ and by a particular choice of
the weights:
\begin{equation}
\label{gold1}
  \pi_j = \delta_j^{D}, \;\; j=1,2
\end{equation}
where $D$ is the (constant) value 
$\frac{\log 2}{\log 5 - \log 2}$. 
This choice originates what is called a {\em uniform
Gibbs measure}.
The first of such I.F.S. is that of Figs. 1 to 3, 
and $D$ is its fractal dimension. 
Indeed, all I.F.S. with the property (\ref{gold1})
are characterized by
the same flat thermodynamic function $D_q=D$.
Clearly, because of eq. (\ref{gold1}), and because
$\pi_1+\pi_2=1$, only one parameter among the map 
weights and contraction rates is left free. By
varying this parameter we can construct
different I.F.S. measures, with the same flat thermodynamics.
What are then the corresponding
dynamical exponents $\beta(\alpha)$ ?
The approximate relation (\ref{multap}) predicts
$\beta(\alpha) \simeq D$ for all $\alpha$.

In Fig. 5  we have considered:
{\bf a:} The I.F.S. with
$\delta_1=\delta_2=\frac{2}{5}$, 
$\pi_1=\pi_2=\frac{1}{2}$, which 
is a ``pure'' Cantor Set.
{\bf b:} The I.F.S. with
$\delta_1=.5090$,
$\delta_2=.2978$, and
$\pi_1 = \frac{3}{5}$.
{\bf c:} The I.F.S. with
$\delta_1=.5293$,
$\delta_2=.2802$, and
$\pi_1 = .6180$.
{\bf d:} The I.F.S. with
$\delta_1=.6033$,
$\delta_2=.2196$, and
$\pi_1 = .6823$.
The first observation we can draw from this
figure is that $\beta$ is not flat,
as shown in \cite{parl}, even if the {\em intermittency range}
in the $[0,5]$ interval is very narrow.
The second, is that the prediction $\beta = D = .7565$ is correct
within two percent at $\alpha=0$ and about five percent 
at $\alpha=5$.
The third, and most important, that
the scaling function $\beta$ is roughly invariant from
case to case.

These results are intriguing: 
the coincidence of the curves in Fig. 5 suggests that
the spectrum of generalized dimensions $D_q$
must play some r\^ole in determining $\beta(\alpha)$:
the fractal measures {\bf a} -- {\bf d} seem to have little
in common beyond having the same flat thermodynamics.
Nevertheless, precisely because in these cases $D_q$ is flat, 
neither 
eq. (\ref{multap}), nor any general relation
of the kind
$\beta(\alpha) = D_{q(\alpha)}$, with $q$ an as yet
unknown function of $\alpha$ can hold rigorously. 

Let us now come to a favourable example:
we can construct a class of measures
for which the renormalization eq. (\ref{est1}) is
{\em exact}: these are the equilibrium measures of the
Julia sets generated by the polynomials 
\begin{equation}
\label{ju1}
  P(z) = z^2 - \lambda,
\end{equation}
where $\lambda \geq 2$ is a real constant.
As we have already remarked,
the Jacobi matrices for these problems can be constructed by
a stable recursion algorithm \cite{danbel}, \cite{barn}.
Non-linearity of the I.F.S. maps stemming from 
eq. (\ref{ju1}) as inverse branches of
$P(z)$ can be treated by considering 
sufficiently high iterations $P^{(l)}$, and a 
theory perfectly analogous to (\ref{sca1}-\ref{multap}) can be
carried out, with the same result.

In fig. 6 we make the usual comparison between the moment 
scaling function, $\beta(\alpha)$, and thermodynamics
\cite{turcb}, $D_{1-\alpha}$:
the curves coincide within numerical precision!
Therefore, one can conclude that discrepancies from 
eq. (\ref{multap}) are due to the non-exactness of eq. (\ref{est1})
when a spectral measure is approximated by I.F.S.,
except for the case of Julia measures, which are 
known to have strong algebraic properties.

Incidentally, we note that for Julia sets, the invariant measure 
coincides with the measure of the asymptotic distribution of 
the zeros of the associated orthogonal polynomials,
the latter being also the physicists' global density of states.
Might it be that the correct quantity entering eq. (\ref{multap})
is this second measure ? The analysis of the I.F.S. data
presented here seems to exclude this case, although we 
cannot exclude that this r\^ole is played
by yet another spectral measure still to be determined.

\section{Conclusions}

We have presented a stable algorithm 
for the determination of lattice Hamiltonian operators possessing a
given spectral measure, in the class of linear I.F.S.
This algorithm consists of a recursive
determination of the associated Jacobi matrix, in the framework of 
the theory of orthogonal polynomials.

The Hamiltonian operators determined in this way are characterized
by almost periodic coefficients: since I.F.S. measures approximate
arbitrary well any measure supported on a Cantor set,
this fact might lead to a proof that almost periodicity is always
associated with this kind of spectra.

In a quantum mechanical context, the Jacobi matrices 
studied here can be employed as models of almost-periodic systems:
the dynamical properties of such systems can be studied in
their essence, having extracted the crucial information on
the related spectral measures.
We have shown that connections between spectral properties and dynamics
go far beyond the conventional RAGE theorems:
in particular, delocalization of particle's position along the
lattice basis can be described by a scaling function $\beta$
governing the moments of order $\alpha$ of the position operator.
Non-constancy of this function translates mathematically
the phenomenon of quantum interference.

We have derived an intriguing relation,
$\beta(\alpha) = D_{1-\alpha}$, linking
dynamics and the thermodynamical properties
of the spectral measure: considering the Jacobi matrices
associated with Julia sets we have constructed a family of 
quantum systems for which the relation is exact, and
we have discussed the reasons for the discrepancies 
present in the general case.
We believe that a further refinement of the results presented
in this paper will lead to a profound understanding of the mathematical and
physical properties of almost-periodic quantum systems.

Finally, we remark that the Jacobi Hamiltonians considered
in this paper
are not simple exotic curiosities, but can also describe
time-resolved
energy absorption in externally perturbed
quantum systems, as well as
electron dynamics in solid-state eterostructures like
super-lattices \cite{sup1}, where 
by varying an alloy concentration along a deposition
axis different
spectral structures can be found \cite{sup2}.
Here, our results may become relevant
in several problems, like --for instance--
the design of lasers and radiation detectors.

\newpage

\begin{center}
{\large {\bf Figure Captions}}
\end{center}

Fig. 1. \\
Orthogonal polynomial $p_8(x)$ of the I.F.S. measure
with maps $(\delta_i,\beta_i,\pi_i)$ $=$ 
$(\frac{2}{5},0,\frac{1}{2})$, 
$(\frac{2}{5},\frac{3}{5},\frac{1}{2})$, with a finite-resolution
representation of the support of the measure obtained 
by plotting a large number of points on the attractor.
Because of the finite size of points, this latter appears as 
a sequence of dashes. Only the symmetrical half 
is shown.
In the inset, the beginning of the 
sequence of $r_n$. The vertical scale ranges
from zero to $\frac{1}{2}$. Lines are merely to guide the
eye.

Fig. 2. \\
Discrete Fourier transform of the $r_n$ sequence 
($n=1,\ldots,2^{13}$) associated with
the I.F.S. of Fig. 1. The constant and $\pi$ frequencies
exceed the vertical scale, and are not reported.

Fig. 3. \\
Plot of the
phase $\Phi$ of the discrete Fourier transform of
the sequence $r_n$ associated with the I.F.S. of Fig. 1 and 2,
to show the
$\pi$ discontinuity close to the value of the main peak
of Fig. 2.

Fig. 4. \\
Multi-fractal dimensions $D_0$ (full diamonds)
and $D_1$ (full squares) and dynamical exponents
$\beta(0)$ (open squares) and
$\beta(1)$ (open diamonds) versus contraction 
rate $\delta_1$, for the family of I.F.S. 
described in the text.

Fig. 5. \\
Scaling functions $\beta(\alpha)$ for the four I.F.S.'s
{\bf a} - {\bf d}
described in the text: {\bf a}: circles; 
{\bf b}: squares;  {\bf c}: triangles; {\bf d}: diamonds.

Fig. 6. \\
Scaling function $\beta(\alpha)$ for the Julia set 
measure with $\lambda = 2.2$ (diamonds) and 
thermodynamical dimensions $D_{1-\alpha}$ (crosses).


\begin{thebibliography}{99}
\bibitem{magn} P.G.Harper, {\em Proc.Roy.Soc.Lon}. A{\bf 68}, (1955) 874; 
M.Ya.Az'bel, {\em Sov.Phys.JETP} {\bf 19} (1964) 634; 
D.R.Hofstadter, {\em Phys. Rev. B} {\bf 14} (1976) 2239;
\bibitem{koh} C.Tang and M.Kohmoto, {\em Phys.Rev.} {\bf B 34} (1986) 2041. 
\bibitem{yam} H.Hiramoto and S.Abe, {\em J.Phys.Soc. Japan} 
{\bf 57} (1988) 230;  {\it ibid.,} (1988) 1365.
\bibitem{geis1}  T.Geisel, R.Ketzmerick, and G.Petschel, 
{\em Phys.Rev.Lett}. {\bf 66},1651(1991); {\it ibid.,} {\bf 67} (1991) 3635;
R.Ketzmerick, G.Petschel and T.Geisel, 
{\em Phys.Rev.Lett}. {\bf 69} (1992) 695.
\bibitem{ig1}  I.Guarneri, {\em Europhys.Lett}. {\bf 10,} 95(1989);
{\em ibid.}, {\bf 21}, 729 (1993).
\bibitem{igm}  I.Guarneri and G.Mantica, {\em Ann. Inst. H.
Poincar\'e} {\bf 61} (1994) 369.
\bibitem{gen} R.del Rio, S.Jitomirskaya, N.Makarov and B.Simon, {\it  
Singular Continuous Spectrum is generic}, preprint 1994. 
\bibitem{gio2}  C.R. Handy and G. Mantica, 
{\em Physica}  {\bf D 43}, (1990) 17-36. 
\bibitem{gio}  G. Mantica and A. Sloan, 
{\em Complex Systems} {\bf 3}, (1989) 37-62.
\bibitem{vrs}  E.R. Vrscay and C.J. Roehrig, {\em Iterated Function Systems and
the Inverse Problem of Fractal Construction Using Moments}, 
in {\em Computers and Mathematics}, 
E. Kaltofen and S.M. Watt Eds., Springer (Berlin, 1989).
\bibitem{gio-guzzi}  D. Bessis and G. Mantica, 
{\em Phys. Rev. Lett.} {\bf 66}, (1991), 2939-2942.
\bibitem{hut}  J. Hutchinson, 
{\em Indiana J. Math.}\ {\bf 30} (1981) 713-747.
\bibitem{dia}  P. Diaconis, M. Shahshahani,
{\em Contemporary Mathematics} {\bf 50} (1986) 173-182.
\bibitem{dem}  M.F. Barnsley and S.G. Demko, 
{\em Proc. R. Soc. London}\ {\bf A 399} (1985) 243-275.
\bibitem{ba2}  M.F. Barnsley, {\em Fractals Everywhere}, Academic Press,
(New York 1988).
\bibitem{kac} M. Case and M. Kac, {\em J. Math. Phys.} {\bf 14} 
(1973) 594.
\bibitem{gax} W. Gautschi, 
{\em Math. Comp.} {\bf 24} (1970) 245-260.
\bibitem{steve}  D. Bessis and S. Demko,  
{\em Physica} {\bf D 47} (1991) 427-438.
\bibitem{gaut} W. Gautschi, in {\em Orthogonal Polynomials}, P. Nevai Ed., 
Kluwer (Dordrecht NL 1990), 181-216.
\bibitem{cap}  G. Mantica, {\em A Stieltjes Technique for Computing
Jacobi Matrices Associated With Singular Measures},
to appear in {\em Constructive Approximations}, (1995).
\bibitem{vrs2}  E.R. Vrscay, {\em I.F.S. Theory and Applications
and the Inverse Problem}, in {\em Fractal Geometry 
and Analysis}, J. B\'elair and S. Dubuc Eds., 
Kluwer, (Dordrecht, NL 1992) 405-468.
\bibitem{andrei} G. Mantica and G.A. Mezincescu, in preparation.
\bibitem{danbel} J. Bellissard, D. Bessis, and P. Moussa,
{\em Phys. Rev. Lett.} {\bf 49} (1982) 702-704.
\bibitem{barn} M.F. Barnsley, J.S. Geronimo, and A.N. Harrington,
{\em Proc. Am. Math. Soc.} {\bf 88} \# 4, (1983) 625-630.
\bibitem{danbak} G.A. Baker, D. Bessis, and P. Moussa,
{\em Physica A} {\bf 124} (1984) 61-77.
\bibitem{dan}  D. Bessis and G. Mantica, 
{\em J. Comp. Appl. Math.} {\bf 48} (1993) 17-32.
\bibitem{parl}  I.Guarneri and G.Mantica, {\em Phys. Rev. Lett.} 
{\bf 73} (1994) 3379.
\bibitem{turca}  D. Bessis, J.D. Fournier, G. Servizi, G. Turchetti,
and S. Vaienti, {\em Phys. Rev.} {\bf A 36}, 920-928 (1987).
\bibitem{turcb}  G. Servizi, G. Turchetti,
and S. Vaienti, {\em Nuovo Cim.} {\bf 101 B}, 285-307 (1988).
\bibitem{sup1} {\em Heterojunction band Discontinuities: Physics and
Device Applications}, F. Capasso and G. Margaritondo Eds.,
Elsevier, B.V. (1987).
\bibitem{sup2} G. Mantica and S. Mantica, {\em Phys. Rev.} {\bf B 46},
7037-7045 (1992).

\end{thebibliography}
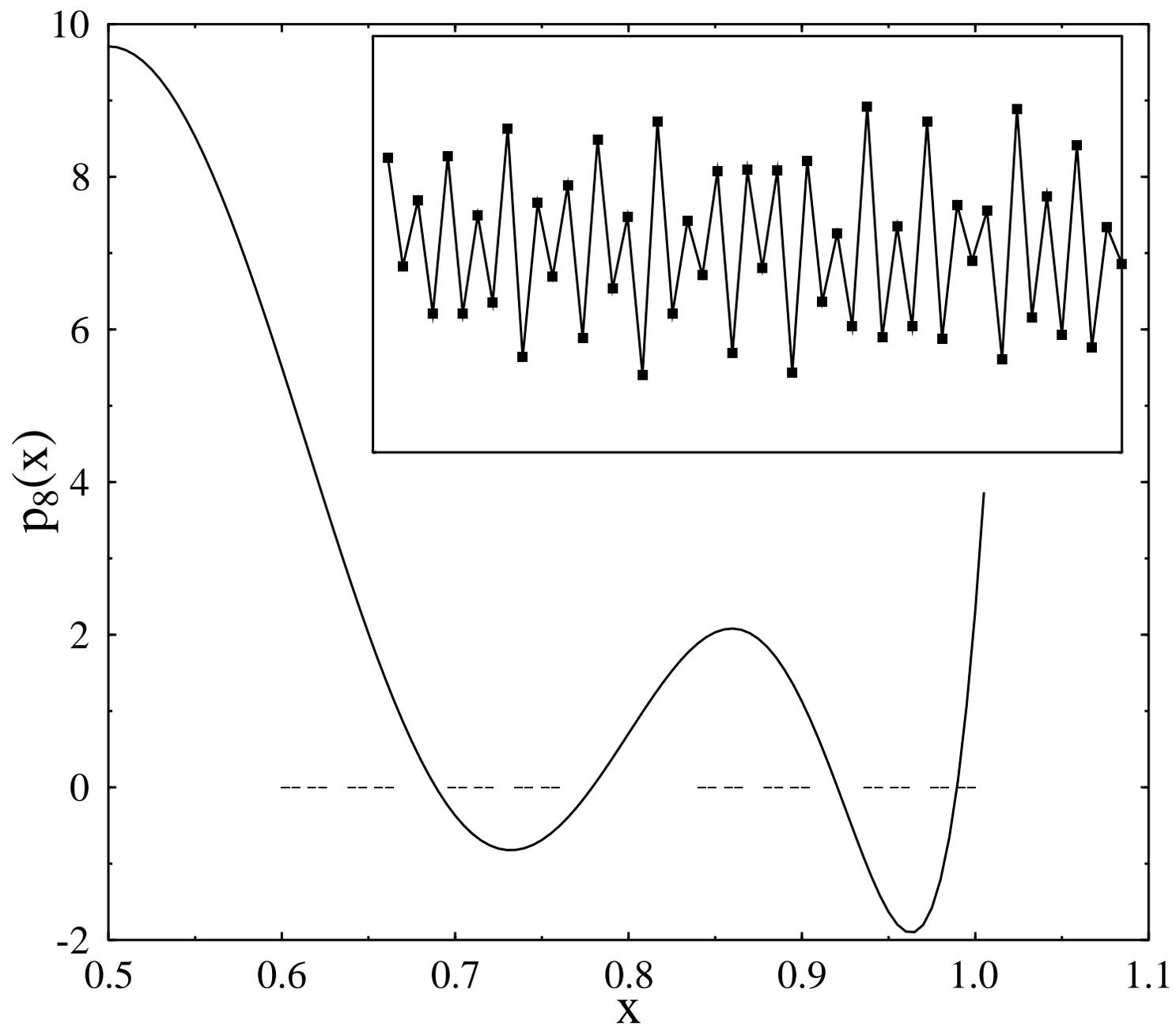
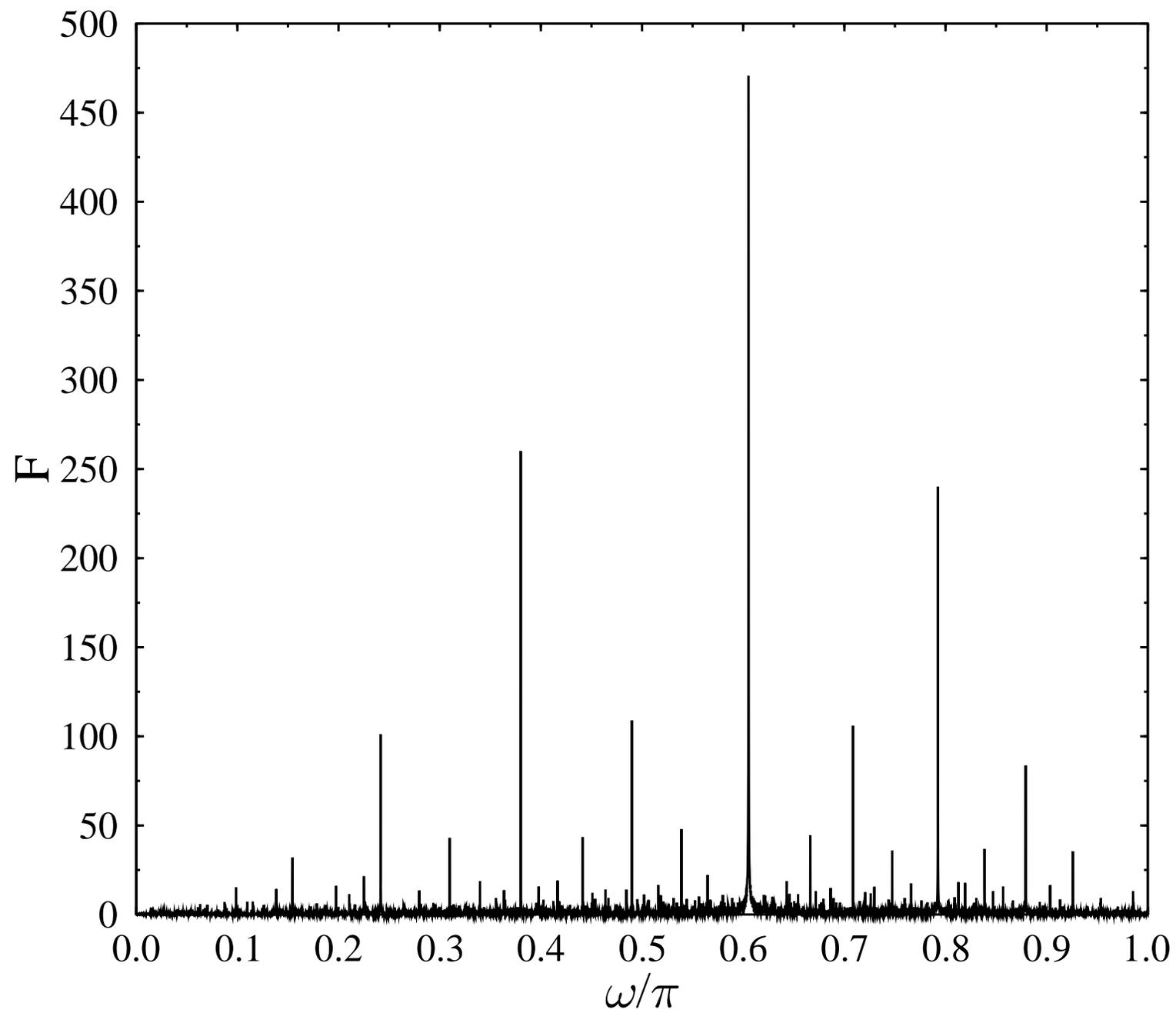
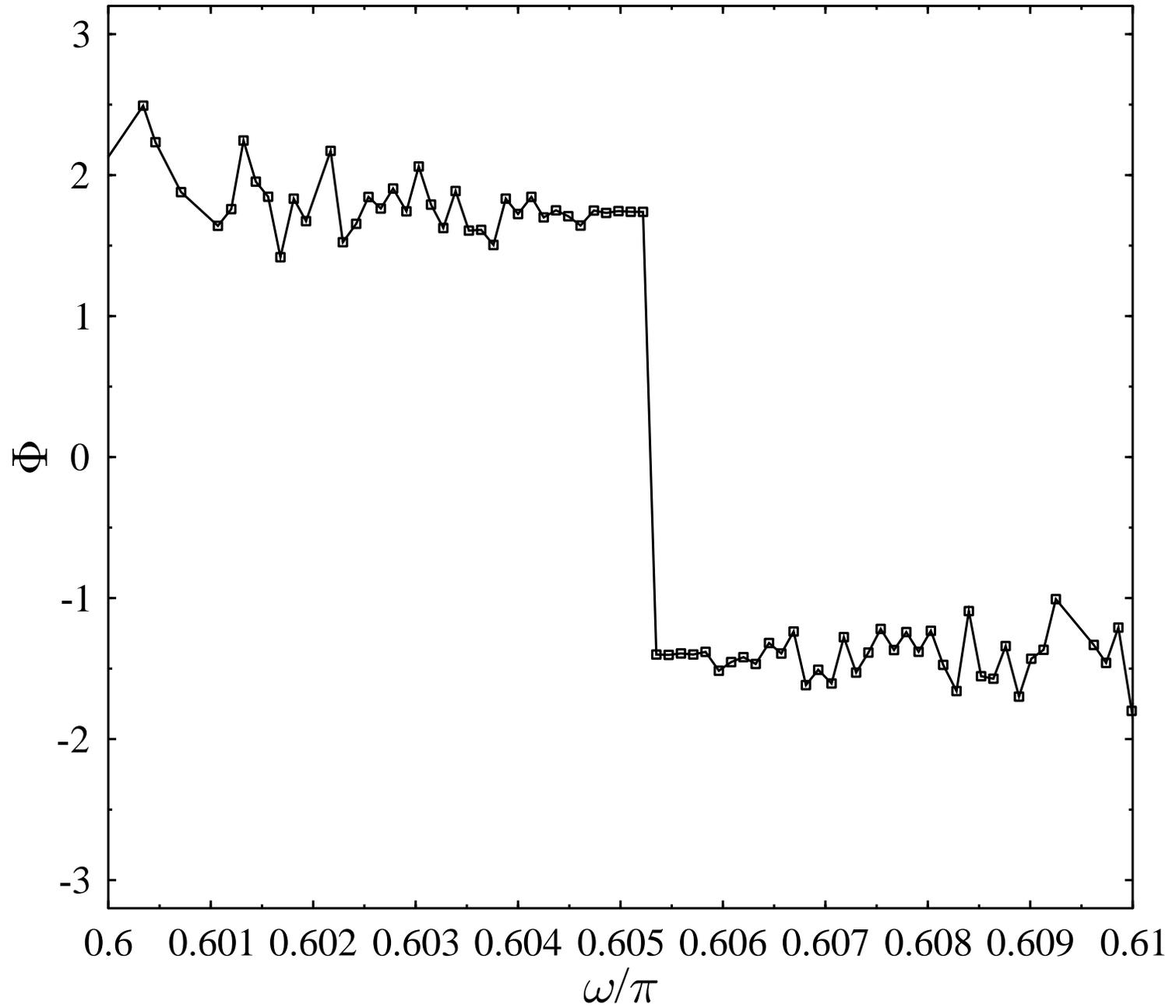
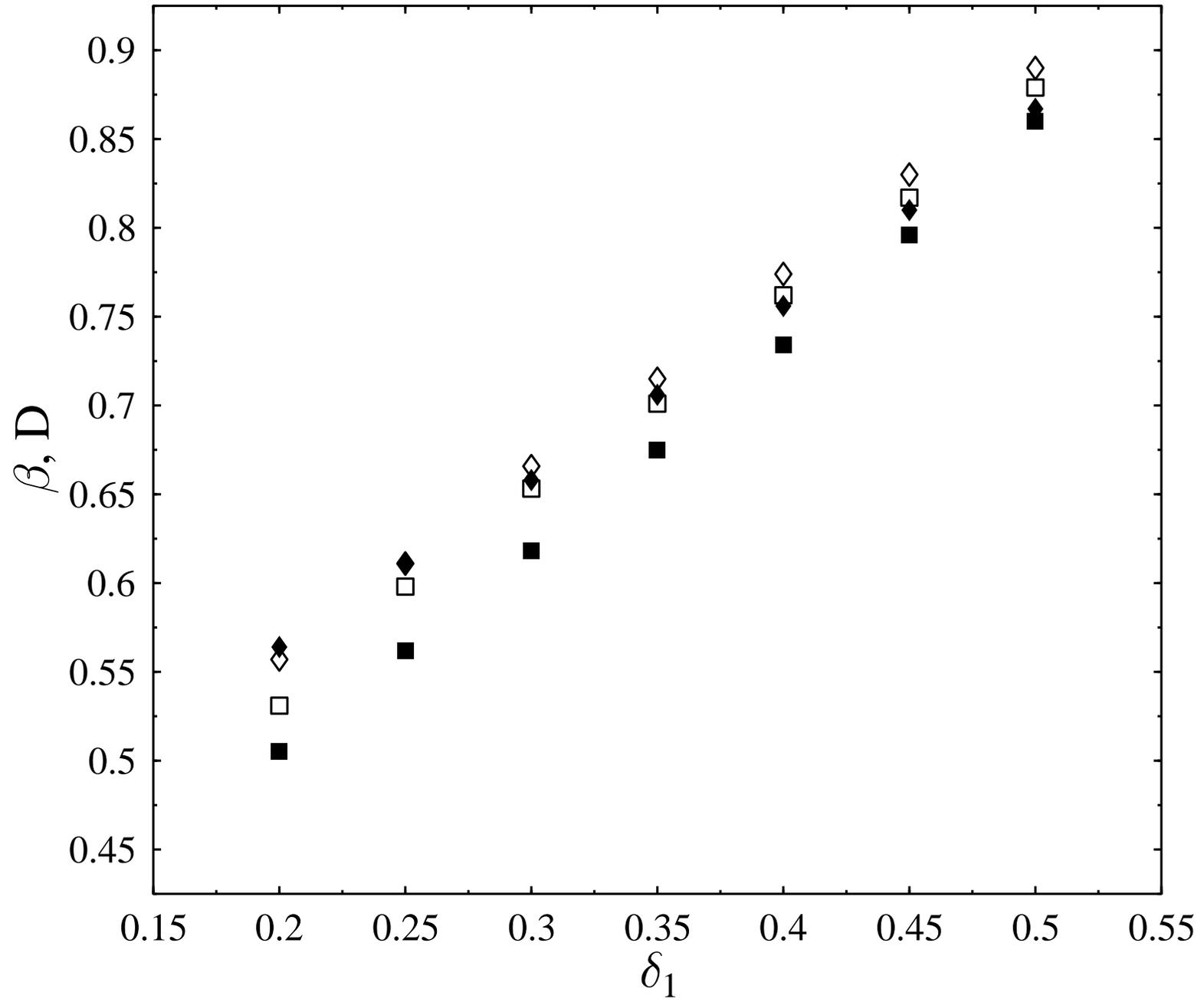
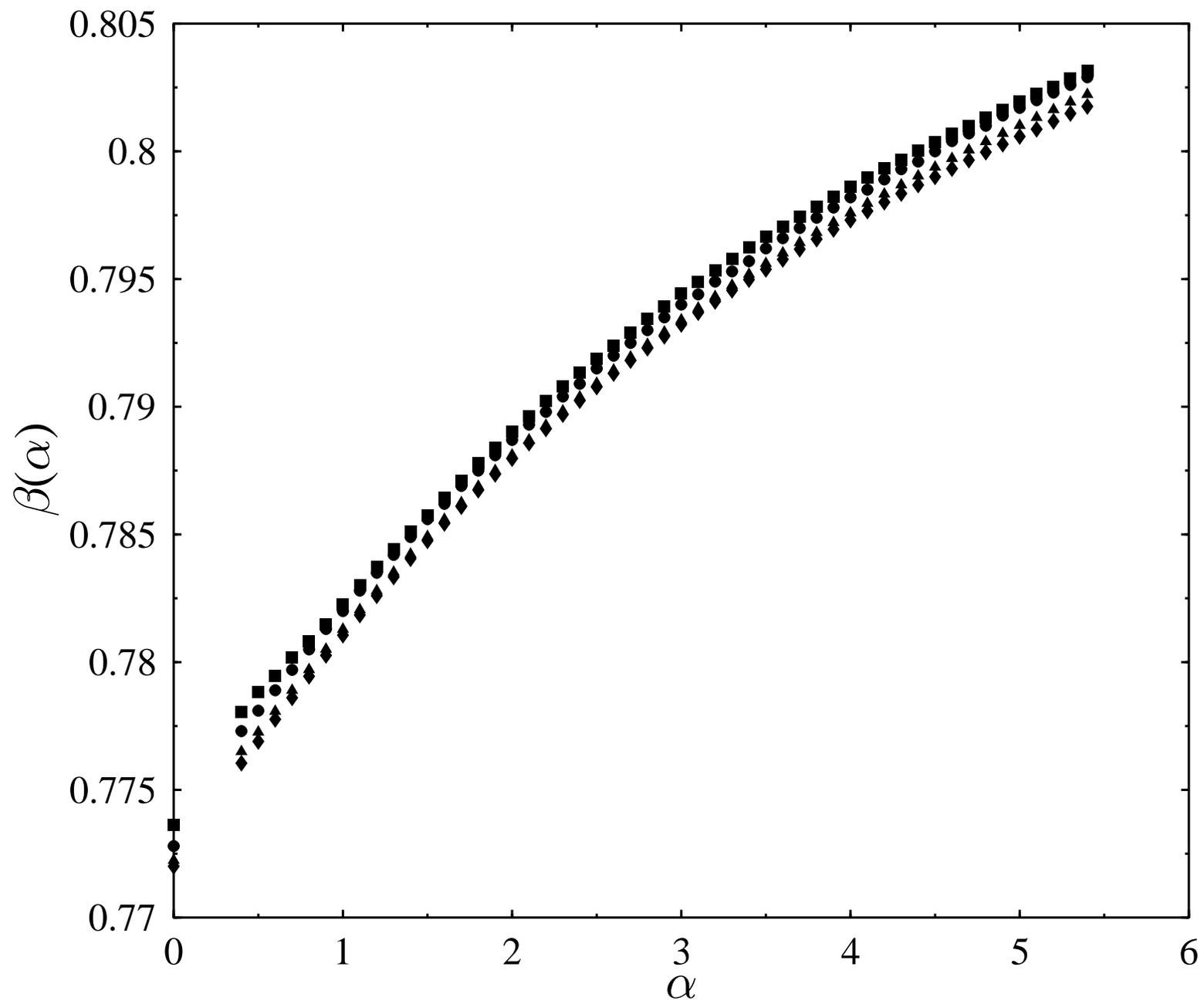
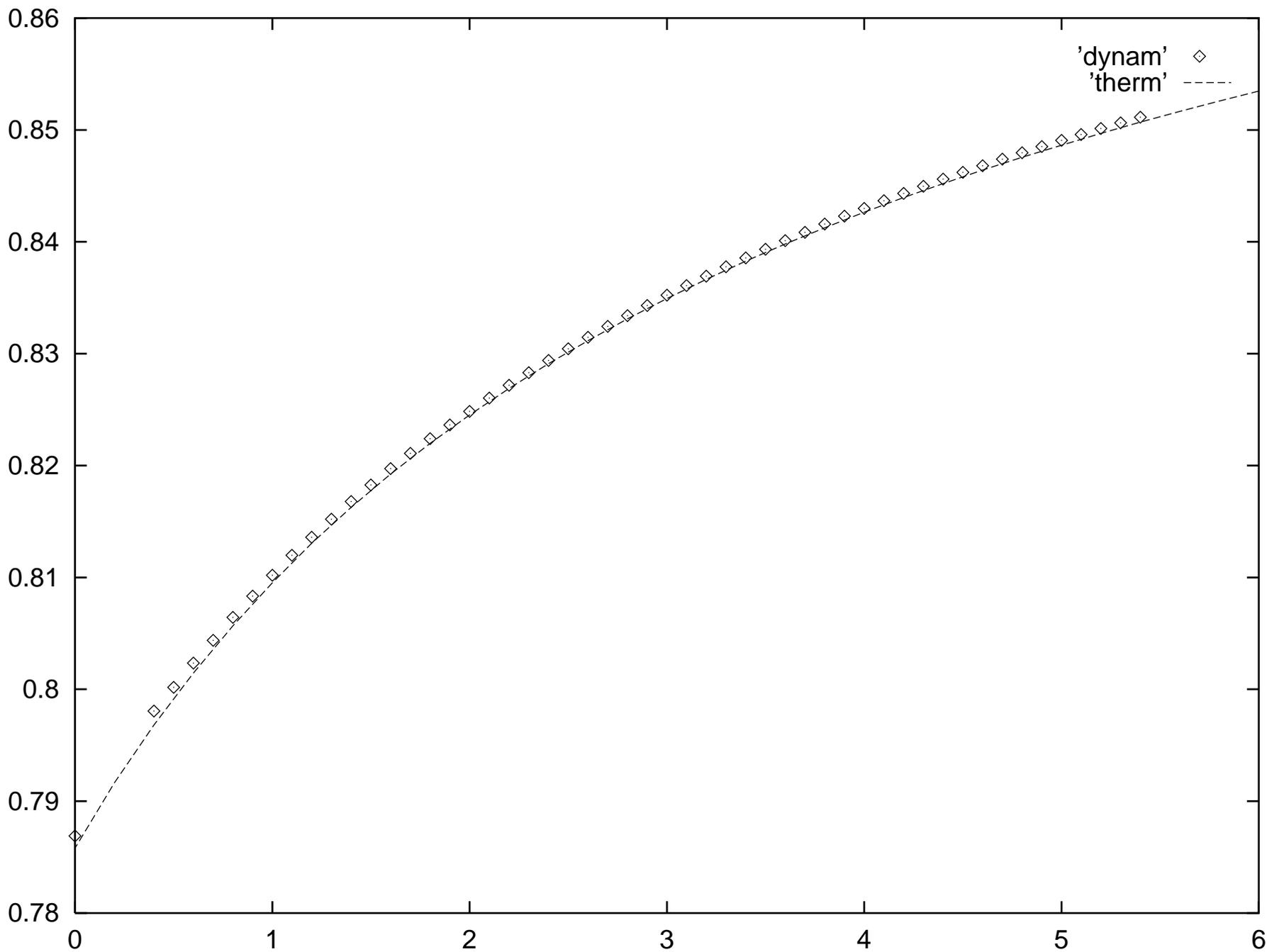
\end{document}